\documentclass[aip,jcp,reprint,noshowkeys]{revtex4-1}
\usepackage{graphicx,dcolumn,bm,xcolor,microtype,multirow,upgreek,amscd,amsmath,amssymb,amsfonts,physics,dcolumn,gensymb}
\usepackage[version=4]{mhchem}

\usepackage{natbib}
\AtBeginDocument{\nocite{achemso-control}}

\usepackage{hyperref}
\hypersetup{
    colorlinks=true,
    linkcolor=blue,
    filecolor=blue,      
    urlcolor=blue,	
    citecolor=blue
}

\usepackage{libertine}
\usepackage[normalem]{ulem}
\usepackage{wrapfig}

\usepackage{xr}
\definecolor{hughgreen}{HTML}{009900}
\definecolor{alexred}{HTML}{B21010}

\usepackage{mathtools}

\newcommand{\mathbold}[1]{\mathbf{#1}}
\newcommand{\SI}{\textcolor{black}{Supporting Information}}

\usepackage{caption} 
\usepackage{float} 

\newcommand{\Eh}{\mathrm{E_h}}

\newcolumntype{d}[1]{D{.}{.}{#1}}
\newcommand{\head}[1]{\multicolumn{1}{c}{#1}}
\newcommand\Tstrut{\rule{0pt}{2.5ex}}         
\newcommand\Bstrut{\rule[-1.5ex]{0pt}{0pt}}   

\newcommand{\pointgroup}[2]{\mathcal{#1}_{\text{#2}}}
\newcommand{\DIVh}{$\pointgroup{D}{4h}$}
\newcommand{\DIIh}{$\pointgroup{D}{2h}$}

\newcommand{\irrep}[2]{\mathrm{#1_{#2}}}
\newcommand{\term}[3]{^{#1}\irrep{#2}{#3}}

\newcommand{\DisAngle}{\phi_{\text{D}}}

\newcommand{\up}[1]{\ ^{#1}}
\newcommand{\NOdet}[1]{^{#1}\Phi}
\newcommand{\kNOdet}[1]{\ket{\NOdet{#1}}}
\newcommand{\NOcoeff}[1]{a_{#1}}
\newcommand{\MO}[2]{^{#1}\phi_{#2}}
\newcommand{\kMO}[2]{\ket{\MO{#1}{#2}}}
\newcommand{\AO}[1]{\chi_{#1}}

\newcommand{\kAO}[1]{\ket{\AO{#1}}}
\newcommand{\JK}[4]{\langle #1 #2 || #3 #4 \rangle}

\newcommand{\sig}{\upsigma}
\newcommand{\sigg}{\sig_\text{g}}
\newcommand{\sigu}{\sig_\text{u}}
\newcommand{\phia}{\phi_{\alpha}}
\newcommand{\phib}{\phi_{\beta}}

\newcommand{\Or}{\mathcal{O}}
\newcommand{\tstat}{\theta_{\text{c}}}
\newcommand{\Ne}{N}
\newcommand{\Nbas}{n}
\newcommand{\Ndet}{m}
\newcommand{\Vnuc}{V_{\text{N}}}

\newcommand{\bD}{\mathbold{P}}
\newcommand{\bF}{\mathbold{F}}
\newcommand{\br}{\mathbold{r}}

\newcommand{\hH}{\hat{H}}
\newcommand{\hh}{\hat{h}}

\renewcommand{\braket}[3]{\langle #1 | #2 | #3 \rangle}
\newcommand{\brkt}[2]{\langle #1 | #2 \rangle}
\newcommand{\hE}{E}

\newcommand{\ie}{i.e.}
\newcommand{\eg}{e.g.}
\newcommand{\etal}{\textit{et al}}
\newcommand{\cf}{\textit{cf.}}
\newcommand{\aufbau}{\textit{Aufbau}}

\newcommand{\qchem}{{\scshape Q-Chem 5.2}}
\newcommand{\libnoci}{{\scshape LIBNOCI}}
\newcommand{\orca}{{\scshape ORCA}}
\newcommand{\mrcc}{{\scshape MRCC}}
\newcommand{\qp}{{\scshape QUANTUM PACKAGE 2.0}}

\newcommand{\Etot}{E_{\mathrm{tot}}}
\newcommand{\RHF}[1]{RHF~#1}
\newcommand{\UHF}[1]{UHF~#1}
\newcommand{\CASSCF}[2]{CASSCF~$(#1,#2)$}

\begin{document} 
\title{A General Approach for Multireference Ground and Excited States using Non-Orthogonal Configuration Interaction}

\author{Hugh~G.~A.~Burton}
\email{hb407@cam.ac.uk}
\affiliation{Department of Chemistry, Lensfield Road, Cambridge, CB2 1EW, U.K.}
\author{Alex~J.~W.~Thom}
\affiliation{Department of Chemistry, Lensfield Road, Cambridge, CB2 1EW, U.K.}

\date{\today}

\begin{abstract}
\begin{wrapfigure}[12]{o}[-1.2cm]{0.4\linewidth}
     \centering
     \includegraphics[width=\linewidth]{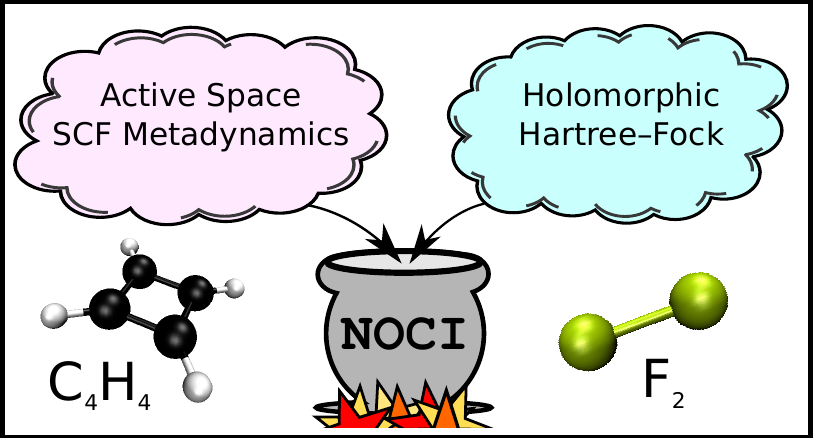}
\end{wrapfigure}
A balanced description of ground and excited states is essential for the description of many chemical processes.
However, few methods can handle cases where static correlation is present, and often these scale very unfavourably with system size.
Recently, multiple Hartree--Fock (HF) solutions have been proposed as a basis for non-orthogonal configuration interaction (NOCI) to provide multireference ground and excited state energies, although applications across multiple geometries have been limited by the coalescence of HF solutions.
Holomorphic HF (h-HF) theory allows solutions to be analytically continued beyond the  Coulson--Fischer points at which they vanish but, until now, this has only been demonstrated for small model systems.
In this work, we propose a general protocol for computing NOCI ground and excited state energies using multiple HF solutions.
To do so, we outline an active space variation of SCF metadynamics that allows a chemically relevant set of HF states to be identified, and describe how these states can be routinely traced across all molecular geometries by exploiting the topology of h-HF solutions in the complex plane.
Finally, we illustrate our approach using the dissociation of the fluorine dimer and the pseudo-Jahn--Teller distortion of cyclobutadiene, demonstrating its applicability for multireference ground and excited states.
\end{abstract}
\maketitle

\section{Introduction:}
A balanced treatment of ground and excited states is essential for the description of a wide range of physical processes, including singlet fission,\cite{Coto2015, Mayhall2016a, Zimmerman2010} electron transfer,\cite{Jensen2018} and primary mechanisms of vision.\cite{Barca2018, Gozem2014, Gozem2012}
However, the presence of strong static correlation in many of these systems --- where the single determinant self-consistent field (SCF) description breaks down --- presents a challenge for many conventional electronic structure techniques.
In principle, full configuration interaction (FCI) can be used to compute the exact energy spectrum for a given basis set.
Practically, however, the exponential scaling of FCI limits its applicability to a minority of relatively small chemical systems.

To circumvent the exponential scaling of FCI, numerous approaches have been developed that provide approximate ground state energies.
In cases dominated by a single reference determinant, these methods range from the mean-field Hartree--Fock (HF) approximation,\cite{SzaboOstlund} to truncated Configuration Interaction (CI), Coupled Cluster (CC) and Perturbation Theory (PT) approaches.\cite{Helgaker, BartlettShavitt}
Meanwhile, for excited states the most common single-reference techniques include Time-Dependent Density Functional Theory (TD-DFT),\cite{Runge1984} uncorrelated CI singles (CIS) \cite{Foresman1992} and Equation of Motion CC (EOM-CC).\cite{Stanton1993, Watts1994a}
Of these, EOM-CC provides the most accurate excitation energies, but is usually limited to only single and double excitations (EOM-CCSD) by the prohibitive cost of including full triple excitations.\cite{Bartlett1995a}

In contrast, the range of techniques available for computing ground and excited states in systems with strong multireference character remains much more limited.
Multi-configurational SCF\cite{Helgaker} (MCSCF) methods provide the prevailing family of techniques, with the Complete Active Space SCF (CASSCF) method being the most widely used approach.\cite{Malmqvist1989} 
In CASSCF, an FCI expansion is solved within a pre-defined active orbital space while the reference orbitals are simultaneously optimised.
To compute multiple excited states and prevent variational collapse, CASSCF must usually be applied using a state-averaged formalism (sa-CASSCF) where the weighted energy of multiple states is optimised rather than a single target state.\cite{Werner1981}
Ultimately, however, CASSCF scales exponentially with the size of the active space and remains a challenge for larger systems, despite recent advances using stochastic\cite{Manni2016} and selected CI (sCI)\cite{Smith2017} approaches.

In addition to conventional techniques, the multiple mean-field solutions to the HF equations have themselves been proposed as approximations to physical excited states.\cite{Barca2018,Gilbert2008a,Besley2009a} 
As a non-linear method, HF theory can yield a large number of multiple solutions\cite{Fukutome1975, Stanton1968, Burton2018} each representing a Slater determinant built from a bespoke set of optimal HF molecular orbitals.
Since the development of SCF metadynamics\cite{Thom2008} and the Maximum Overlap Method (MOM),\cite{Besley2009a} the identification of multiple SCF solutions has become relatively routine. 
In many cases, however, multiple HF states can can break the symmetries of the true Hamiltonian, leading to wave functions that do not satisfy the molecular point group symmetry, or are not eigenfunctions of the spin operators $\hat{\mathcal{S}}^2 $ or $ \hat{\mathcal{S}}_{\text{z}}$.\cite{StuberPaldus}

At the HF level, the onset of strong static correlation is indicated by the break-down of the single determinant approximation.
The restricted HF (RHF) ground state in \ce{H2}, for example, overestimates the binding energy by incorrectly dissociating into a linear sum of the physical ``radical'' and ``ionic'' states.\cite{Coulson1949}
In contrast, the unrestricted HF (UHF) approach provides the correct behaviour at dissociation by allowing electrons of opposite spin to occupy different spatials orbitals and localise on opposing hydrogen atoms.\cite{SzaboOstlund}
Moreover, the signature of multireference character --- namely several significant reference determinants --- manifests directly in the degeneracy of the HF solutions.

To exploit the natural description of static correlation provided by multiple HF solutions, recent research has focussed on using these states as a multireference basis for CI calculations.\cite{Thom2009, Sundstrom2014, Yost2013, Mayhall2014, Burton2016, Oosterbaan2018a, Jensen2018}
Since multiple HF solutions are not required to be mutually orthogonal, this CI takes the form of a non-orthogonal CI (NOCI).\cite{Thom2009}
By constructing a wave function as a linear combination of non-orthogonal HF solutions, NOCI is able to capture electron correlation with a scaling of $\Or( \Ndet^2\ \text{max}(\Ne^3, \Nbas^2 ))$, where $\Ndet$ is the number of determinants used in the expansion, $\Ne$ is the number of electrons, and $\Nbas$ is the number of basis functions.\cite{Thom2009} 
In addition, including all degenerate symmetry-broken HF states in the NOCI basis enables the restoration of broken symmetries in a similar style to Projected HF (PHF)\cite{JiminezHoyos2012, Lowdin1955, Scuseria2011} and Half-Projected HF (HPHF)\cite{Smeyers1973, Smeyers1974, Cox1976} approaches.
However, in contrast to the variation-after-projection style of PHF methods, NOCI provides a projection-after-variation formalism where symmetry is partially restored in a single CI expansion.
As a result, NOCI provides only approximate symmetry restoration for infinite symmetry groups (\eg{} spin), but avoids the challenging non-linearity and self-consistency of PHF methods.

As an inherently multireference approach, NOCI is well-suited to capturing static correlation and can be made size-consistent by ensuring a suitable set of determinants is used.\cite{Sundstrom2014}
Since determinants are constructed from different sets of orbitals --- each optimised individually at the SCF level --- NOCI can provide a more balanced treatment of ground and excited states, leading to its application for multi-electron excitations,\cite{Sundstrom2014} core excitations,\cite{Oosterbaan2018a,Oosterbaan2019} and charge transfer processes.\cite{Jensen2018}
Moreover, the non-orthogonality of the NOCI basis can lead to more efficient and compact multideterminantal expansions that provide useful initial guess orbitals for active space techniques,\cite{Krausbeck2014} or trial nodal surfaces for quantum Monte-Carlo techniques.\cite{Pathak2018, LandinezBorda2019}
The combination of relatively low-order polynomial scaling, compact multideterminant expansions, and the ability to ensure size-consistency presents NOCI as a promising alternative to CASSCF for treating multireference ground and excited states.

Despite the progress of NOCI methods, their application using symmetry-broken HF states over a range of molecular geometries has, until recently, been limited by the disappearance of HF solutions at Coulson--Fischer points.\cite{Coulson1949}
At such points, there is a sudden reduction in the size of the NOCI basis set leading to discontinuities in the NOCI energy.
To prevent these discontinuities, the holomorphic Hartree--Fock (h-HF) approach has been developed as a method for analytically continuing real HF solutions beyond the points at which they vanish.\cite{Hiscock2014, Burton2016, Burton2018}.
In h-HF theory, a new energy function is defined by removing the complex-conjugation of orbital coefficients from the conventional HF equations.
This transformation can be seen as a complex analytic continuation of real HF theory and results in a non-Hermitian theory in which the energy is a complex-analytic polynomial of the orbital coefficients.
Significantly, the stationary points of the h-HF energy appear to exist across all molecular geometries,\cite{Burton2018} extending with complex orbital coefficients beyond the Coulson--Fischer points where their real counterparts vanish.
As a result, h-HF stationary states provide a continuous basis for NOCI calculations and yield continuous potential energy surfaces across all geometries.\cite{Burton2016, Burton2018}

Although the combination of h-HF and NOCI has been shown to provide accurate approximations to FCI for small systems,\cite{Burton2016, Burton2018} its applicability in more chemically relevant examples is yet to be demonstrated and some practical issues have persisted.  
In particular, since (in the absence of a magnetic field) the h-HF energy is symmetric along the real orbital coefficient axis, it can be difficult to identify the correct complex direction in which to follow h-HF solutions when their real counterparts vanish.
Moreover, as the SCF solution space can become very large,\cite{Burton2018} identifying a suitable basis set of chemically relevant SCF states can prove difficult.

In the current Paper we outline a general approach for applying NOCI in combination with h-HF theory.
We describe how, by applying SCF metadynamics in an active orbital space, one can identify a suitable set of chemically relevant determinants.
We then demonstrate a simple approach that allows h-HF solutions to be followed around a Coulson--Fischer point and into the complex orbital coefficient plane.
Finally, we illustrate our general approach by considering to the dissociation of \ce{F2} and the pseudo-Jahn--Teller distortion of cyclobutadiene.

\section{Theory}
\subsection{Non-Orthogonal Configuration Interaction}
The NOCI wave function is constructed as a linear combination of $\Ndet$ mutually non-orthogonal basis states $\{ \kNOdet{x} \}$ as
\begin{equation}
\ket{\Psi} = \sum_{x}^{\Ndet} \kNOdet{x} \NOcoeff{x},
\label{eq:NOCIWavefunction}
\end{equation}
where we employ the non-orthogonal tensor notation of Head-Gordon \etal{}.\cite{Head-Gordon1998}
Each state $\kNOdet{x}$ corresponds to a single Slater determinant constructed from a bespoke set of $\Ne$ occupied molecular orbitals (MOs), $\{ \kMO{x}{i} \}$, which themselves are formed from a linear combination of $\Nbas$ (non-orthogonal) atomic orbitals (AOs), $\{ \kAO{\mu} \}$, as
\begin{equation}
\kMO{x}{i} = \sum_{\mu}^{\Nbas} \up{x}C^{\mu \cdot}_{\cdot i} \kAO{\mu}.
\end{equation}
The NOCI eigenstates are identified by solving the generalised eigenvalue problem
\begin{equation}
\sum_{x}^{\Ndet} \left( H^{wx} - E S^{wx} \right) \NOcoeff{x} = 0,
\label{eq:NOCIGenEigval}
\end{equation}
where $H^{wx} = \braket{\NOdet{w}}{\hH}{\NOdet{x}}$ and $S^{wx} = \brkt{\NOdet{w}}{\NOdet{x}}$ are the Hamiltonian and overlap matrix elements in the non-orthogonal basis.
Following the approach detailed in Ref.~\onlinecite{Thom2009}, $H^{wx}$ and $S^{wx}$ are computed by constructing a biorthogonal set of orbitals using L\"{o}wdin's pairing approach\cite{Lowdin1962, Amos1961} and then applying the generalised Slater--Condon rules.\cite{MayerBook}
However, for complex-valued MOs we note that the correct form of the unweighted codensity matrices outlined in Ref.~\onlinecite{Thom2009} is 
\begin{equation}
\up{wx}P^{\mu \nu}_{i} = (\up{x}\widetilde{C})^{\mu \cdot}_{\cdot i} (^{w}\widetilde{C}^{*})^{\cdot \nu}_{i \cdot},
\label{eq:ComplexCodensity}
\end{equation}
in contrast to Eq.~(3) provided therein.

In comparison to orthogonal CI, selecting a relevant set of determinants to form the NOCI basis is less trivial.
In particular, the lack of a universal set of MOs across the non-orthogonal determinants removes the intuitive concept of excitation levels (\ie{} singles, doubles) that pervades orthogonal approaches.
Moreover, the lack of orthogonality between the MOs of multiple non-orthogonal determinants allows excitations of any order to couple in the NOCI expansion.
As a result, more Hamiltonian and overlap matrix elements usually need explicit computation, although  non-orthogonality can also lead to shorter multi-determinant expansions.

A number of approaches have been proposed for constructing non-orthogonal basis sets, ranging from the use of SCF metadynamics to identify multiple HF solutions,\cite{Thom2008} to spin-flip\cite{Mayhall2014} (SF-NOCI) and CIS\cite{Oosterbaan2018a,Oosterbaan2019} (NOCIS) inspired approaches.
In SF-NOCI, non-orthogonal determinants are constructed by independently relaxing all possible spin-flip excited determinants from a restricted high-spin reference using a frozen active space SCF procedure.\cite{Mayhall2014}
In contrast, the NOCIS approach builds the non-orthogonal basis by first individually optimising a set of restricted high-spin reference determinants, each corresponding to the removal of a different core electron, and then reattaching the excited electron to all possible virtual orbitals.\cite{Oosterbaan2018a}

Significantly, in both SF-NOCI and NOCIS, the basis states are computed by partially optimising multiple determinants constructed as excitations from a symmetry pure reference.
In constrast, by building the NOCI basis from multiple HF solutions identified using SCF metadynamics, one can supplement symmetry-pure determinants with symmetry-broken HF states and capitalise on the electron correlation symmetry-broken states provide at the mean-field level.
However, to ensure the existence of symmetry-broken HF states across all geometries --- and thus guarantee continuous NOCI energies --- HF solutions must be analytically continued beyond the Coulson--Fischer points at which they vanish.
To do so, we construct the basis for NOCI using the stationary points of the h-HF equations.\cite{Hiscock2014, Burton2016}

\subsection{Holomorphic Hartree--Fock Theory}
Using the h-HF approach, real HF solutions can be analytically continued across all geometries by identifying the stationary points of the h-HF energy function\cite{Hiscock2014}
\begin{equation}
\hE = \frac{\braket{\Phi^{*}}{\hH}{\Phi}}{\brkt{\Phi^{*}}{\Phi}}.
\label{eq:HoloEnergyFunc}
\end{equation} 
Here $\hH$ is the conventional electronic Hamiltonian
\begin{equation}
\hH = \Vnuc + \sum_{i}^{\Ne} \hh \qty(i) + \sum_{i<j}^{\Ne} \frac{1}{r_{ij}},
\end{equation}
where $\Vnuc$ is the nuclear repulsion, $\hh \qty(i)$ are the one-electron operators and $r_{ij} = | \br_{i} - \br_{j} |$ defines the distance between electrons $i$ and $j$.
In constrast to the complex-Hermitian extension of HF,\cite{Ostlund1972} the h-HF equations form a complex-analytic continuation of real HF theory by allowing the MO coefficients to become complex without introducing complex conjugation into the energy function.
As a result, we find a constant number of h-HF stationary points to Eq.~\eqref{eq:HoloEnergyFunc} across all geometries and, where real HF solutions vanish, their h-HF counterparts extend with complex MO coefficients.\cite{Hiscock2014, Burton2018}
In turn, the h-HF solutions provide a continuous basis for NOCI.\cite{Burton2016, Burton2018}

Notably, since Eq.~\eqref{eq:HoloEnergyFunc} satisfies the Cauchy--Riemann conditions,\cite{Fischer} it is a complex-analytic function with no concept of minima and maxima.
However, stationary points are still well-defined as points where the magnitude of the gradient becomes zero.
Identifying optimal h-HF states corresponding to the stationary points of Eq.~\eqref{eq:HoloEnergyFunc} requires only minor modifications to the conventional SCF procedure.\cite{Burton2016}
In particular, removing the complex-conjugation of the MO coefficients leads to complex-symmetric (\cf{} Hermitian in conventional complex HF) density, $\bD$, and Fock, $\bF$, matrices defined as
\begin{equation}
P^{\mu \nu} = \sum_{i}^{\Ne} C^{\mu \cdot}_{\cdot i} C^{\cdot \nu}_{i \cdot}
\label{eq:HoloDensityMatrix}
\end{equation}
and 
\begin{equation}
F_{\mu \nu} = h_{\mu \nu} + \sum_{\sigma \tau}^{\Nbas} \JK{\mu}{\sigma}{\nu}{\tau}  P^{\tau \sigma},
\label{eq:HoloFockMatrix}
\end{equation}
where $h_{\mu \nu}$ and $\JK{\mu}{\sigma}{\nu}{\tau}$ are the one-electron and anti-symmetrised two-electron integrals respectively.\cite{SzaboOstlund}
Since the optimal molecular orbitals are now given by the eigenvectors of a complex-symmetric matrix, the MO coefficients must form a complex-orthogonal set\cite{Craven1969} (\cf{} unitary eigenvectors for Hermitian matrices) satisfying
\begin{equation}
\sum_{\mu \nu}^{\Nbas} C^{\cdot \mu}_{i \cdot} S_{\mu \nu}  C^{\nu \cdot}_{\cdot j} = \delta_{ij},
\label{eq:HoloComplexOrth}
\end{equation} 
where $S_{\mu \nu} = \brkt{\AO{\mu}}{\AO{\nu}}$ is the (real) AO overlap matrix.
Finally, the h-HF energy and orbital energies are complex in general and the \aufbau{} ordering principle\cite{SzaboOstlund} cannot be applied to select the occupied orbitals on each SCF iteration.
Instead, we have found a complex-symmetric analogue of MOM to provide an effective alternative approach.\cite{Burton2016}

Since all real HF states are also solutions the h-HF equations, from hereon we use the term h-HF to refer only to the stationary points of Eq.~\eqref{eq:HoloEnergyFunc} with complex orbital coefficients.

\section{A General Approach}
\subsection{Active Space SCF Metadynamics}
\begin{figure*}[tp!]
\includegraphics[width=\textwidth ]{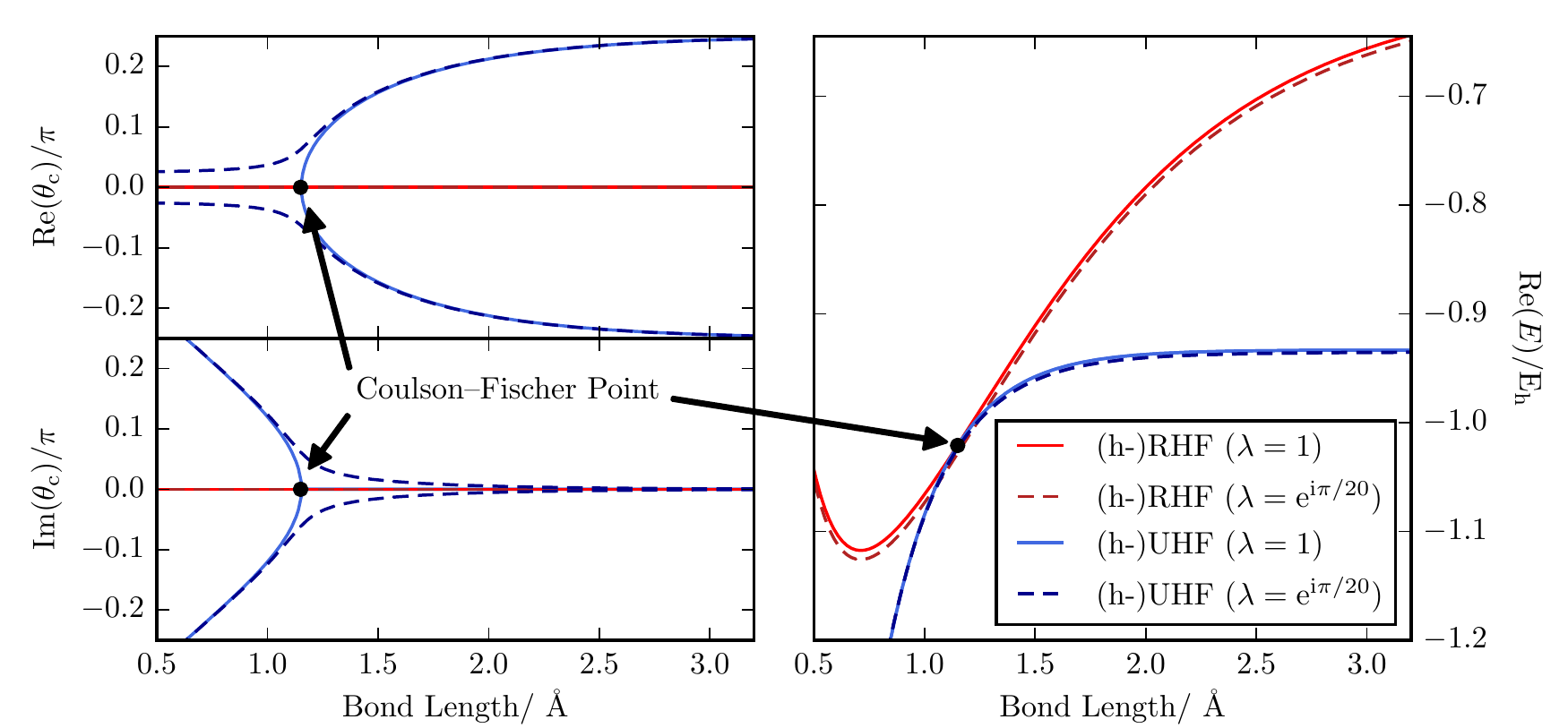}
\caption{The stationary values $\tstat$ (left) for the RHF (red) and h-UHF (blue) states in the unperturbed $\lambda=1$ (solid) and perturbed $\lambda = \text{e}^{\text{i} \pi /20}$ (dashed) cases.
In the unperturbed case, the three states coalesce simultaneously at the Coulson--Fischer point (black dot), while in the perturbed case the h-UHF solution can be followed smoothly into the complex plane.
The h-HF energy (right) is only slightly affected by the perturbation.}
\label{fig:H2PastCFP}
\end{figure*}

Identifying HF states to form a non-orthogonal basis is an integral, yet challenging, component of NOCI calculations.
SCF metadynamics provides an effectively black-box approach for locating multiple stationary points.\cite{Thom2008}
Starting from an optimised solution, SCF metadynamics generates new initial guess determinants by randomly mixing occupied and virtual MOs, and then optimises these determinants in the presence of a biasing potential to prevent re-converging onto a previously known stationary point.
However, even for modest systems, the number of states located may become very large and identifying a suitable set of chemically relevant determinants can be difficult.

Fortunately, in many cases, a dominant subset of ``active'' MOs can be identified that strongly influence the characteristics of the potential energy surface.
The most relevant HF determinants are usually related to different active orbital occupations, along with symmetry-broken states formed from mixing these MOs.
We can therefore define an active space SCF metadynamics approach that uses these active orbitals to identify a suitable subset of HF states as follows.
Starting from an initial symmetry-pure reference determinant, \eg{} the RHF ground state, a metadynamics calculation is run in which only the active MOs are allowed to mix and where the SCF optimisation proceeds only in this active space --- \ie{} the inactive orbitals remain frozen throughout.
This process leads to a set of determinants that differ only the composition of their active MOs, but are not themselves fully optimised HF stationary points.
Subsequently relaxing the inactive orbitals by optimising each determinant in the full orbital space then yields true HF solutions that form the basis for NOCI.

Significantly, the number of partially optimised HF states with frozen inactive orbtials is controlled by the size of the active space and is much smaller than the number of states in the full unfrozen HF space.
As a result, active space SCF metadynamics provides a more manageable approach to identifying a chemically relevant basis of HF states for NOCI.

\subsection{Moving Past the Coulson--Fischer Point}

While h-HF theory allows real HF states to be analytically continued into the complex plane, identifying the correct complex direction to follow solutions in the MO coefficient space has previously proved difficult.
In particular, since the h-HF energy is symmetric about the real MO coefficient axis, SCF calculations starting from a real guess (for example a real HF solution from a previous geometry) show no preference to move towards any particular complex direction.
Additionally, the coalescence of symmetry-broken states often coincides exactly with a real symmetry-pure solution at a cusp catastrophe,\cite{Burton2018} and thus attempts to trace symmetry-broken states into the complex plane often remain stuck on real symmetry-pure solutions instead.

Routinely following h-HF states past the Coulson--Fischer point into the complex requires an understanding of the complex topology of h-HF solutions.
Recently, by scaling the electron-electron interaction using the complex parameter $\lambda$, \ie{} creating the perturbed Hamiltonian
\begin{equation}
\hH_{\lambda} = \Vnuc + \sum_{i}^{\Ne} \hh \qty(i) + \lambda \sum_{i < j}^{\Ne} \frac{1}{r_{ij}},
\end{equation}
we have shown that multiple h-HF states form a continuous interconnected manifold in the complex $\lambda$-plane.\cite{Burton2019b}
Each individual h-HF state is given by a different branch of a Riemann surface, with Coulson--Fischer points forming \textit{isolated} exceptional points corresponding to the branch points of the Riemann surface.\cite{Burton2019b}
Significantly, we can use this continuous topology of h-HF states to follow solutions \textit{around} a Coulson--Fischer point by tracing a suitable $\lambda$-trajectory in the complex plane.

We illustrate this idea by considering the multiple HF solutions of \ce{H2} in a minimal basis set (STO-3G) using two orbitals $\phia$ and $\phib$ parameterised by the complex angle $\theta$,
\begin{subequations}
\begin{align}
	\phia	& = \sigg \cos \theta	+ \sigu \sin \theta,
	\\
	\phib	& = \sigg \cos \theta	- \sigu \sin \theta.
\end{align}
\end{subequations}
Stationary points correspond to the critical values $\tstat$.
Additionally to the RHF $\sigg^2$ state (red solid line in Fig.~\ref{fig:H2PastCFP}), in the unperturbed case ($\lambda = 1$) a doubly degenerate pair of real spatial symmetry-broken UHF (sb-UHF) solutions exist in the dissociation limit (blue solid line in Fig.~\ref{fig:H2PastCFP}).
These correspond to the diradical configurations \ce{^{$\upharpoonleft$}H\bond{...}H^{$\downharpoonright$}} and \ce{^{$\downharpoonleft$}H\bond{...}H^{$\upharpoonright$}}.\cite{Coulson1949, SzaboOstlund, Hiscock2014}
As the bond length is shortened, the sb-UHF solutions coalesce with the RHF state at the Coulson--Fischer point, before continuing into the complex-$\theta$ plane as h-UHF solutions\cite{Hiscock2014} (left panel in Fig.~\ref{fig:H2PastCFP}).

To move smoothly from the real sb-UHF branch, past the Coulson--Fischer point and onto the complex h-UHF branch, we first perturb the real solutions into the complex plane by taking $\lambda = \text{e}^{\text{i} \pi /20}$  (dashed lines in Fig.~\ref{fig:H2PastCFP}). 
Although the $\tstat$ for the RHF state is by fixed spatial symmetry and remains unchanged,  the corresponding $\tstat$ values for the sb-UHF state become complex even for long bond lengths, smoothly connecting the real sb-UHF and complex h-UHF regimes \textit{without ever} passing through the Coulson--Fischer point.
Relaxing these pertubed stationary points back to $\lambda = 1$ at each geometry then leads to the unperturbed states required for NOCI.

\subsection{Combined Protocol}

We are now in a position to define a general protocol that combines active space SCF metadynamics, h-HF and NOCI to compute the ground and excited states of molecular systems.
In most systems, the multiple symmetry-broken HF states of interest occur at geometries where strong static correlation is present, for example dissociaton or transition states. 
Therefore, to ensure every important solution is captured, the initial active space SCF metadynamics calculation is run at all geometries of particular interest.
The states identified at each specific structure can then be connected by tracing solutions across the intermediate geometries, including identifying h-HF extensions for any real solutions that coalesce and vanish.

Our combined approach can be summarised as follows:
\begin{enumerate}
\item{Identify real HF solutions at the geometries of interest using active space SCF metadynamics.}
\item{Perturb states off the real axis using a complex $\lambda$ scaling.}
\item{Trace the perturbed solutions across all geometries, identifying any corresponding complex h-HF solutions required.}
\item{Relax all states back onto the real axis.}
\item{Compute NOCI energies using the resulting basis of multiple h-HF solutions.}
\end{enumerate}
We have implemented this approach, comprising active space SCF metadynamics, h-HF and NOCI, in a new dedicated \libnoci{} library available in \qchem{}.\cite{QCHEM}

\section{Results and Discussion}

We now illustrate our general approach by considering the ground-state dissociation of \ce{F2} and the ground and excited states in the pseudo-Jahn--Teller distortion of cyclobutadiene.
Both systems exhibit a challenging electronic structure including multireference character and strong correlation effects.
To assess the performance of combining h-HF and NOCI, we compare our results to currently available methods including CIS, CASSCF, EOM-CC and FCI.

\subsection{Computational Details}

All h-HF and NOCI energies, along with geometry optimisations (performed at the RHF level) and CIS excitation energies, are calculated using \qchem{}.\cite{QCHEM}
Where h-HF energies become complex, only the real part is plotted.
CASSCF energies are computed using the \orca{} quantum chemistry package\cite{ORCA} and CC calculations, including EOM-CC,\cite{Kallay2004} are run using \mrcc{}.\footnote{{\scshape MRCC}, a quantum chemical program suite written by M. K\'{a}llay, P. R. Nagy, Z. Rolik, D. Mester, G. Samu, J. Csontos, J. Cs\'{o}ka, B. P. Szab\'{o}, L. Gyevi-Nagy, I. Ladj\'{a}nszki, L. Szegedy, B. Lad\'{o}czki, K. Petrov, M. Farkas, P. D. Mezei, and B. H\'{e}gely. See also Z. Rolik, L. Szegedy, I. Ladj\'{a}nszki, B. Lad\'{o}czki, and M. K\'{a}llay, {\it J. Chem. Phys.} {\bf 139}, 094105 (2013), as well as: \texttt{www.mrcc.hu}}
To obtain approximate FCI comparisons for cyclobutadiene, we use a sCI method\cite{Holmes2016, Evangelisti1983, Huron1973, Giner2013} shown to provide near FCI accuracy for both ground and excited states.\cite{Holmes2017b, Loos2018b, Loos2018c, Chien2017a, Scemama2018b}
In particular, we use the CIPSI (CI using a perturbative selection made iteratively) algorithm\cite{Evangelisti1983, Huron1973, Giner2013} implemented in \qp{}\cite{QuantumPackage2.0} using the frozen core approximation.
Futher information on our sCI calculations, including detailed results, is provided in the \SI{}.

The cc-pVDZ basis set is used throughout, and all energies are provided in units of Hartrees, $\Eh$.

\subsection{Dissociation of \ce{F2}}

\begin{figure}[tbp!]
\includegraphics[width=0.5\textwidth, keepaspectratio=true, trim={0.15em 0ex 0.3em 0ex}, clip]{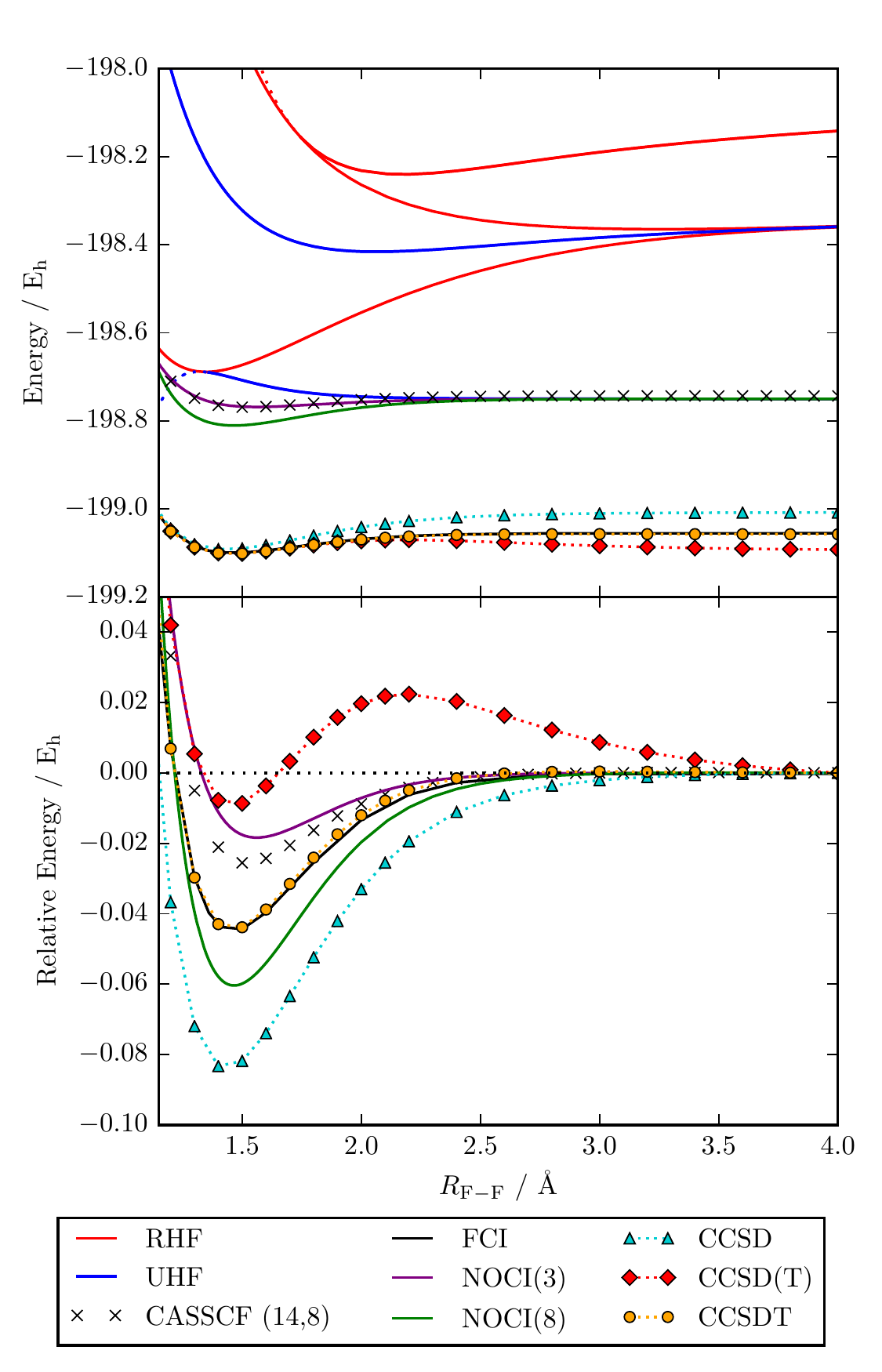}
\caption{Absolute (top) energies for the eight multiple h-HF solutions of \ce{F2} (cc-pvdz), along with NOCI(3) and NOCI(8) ground state energies computed using only the three lowest and all eight h-HF states respectively.
Binding curves (bottom) are computed relative to the value of each method at $R_{\mathrm{F-F}} = 4.0$~\AA{} and compared to FCI energies from Ref.~\onlinecite{Bytautas2010}.}
\label{fig:fluorine_cc-pvdz}
\end{figure}

Due to the combination of strong static and dynamic correlation effects, the ground-state binding curve for \ce{F2} is notoriously difficult to compute.\cite{Sears2003, Yang2013,Gordon1987, Krylov2002, Laidig1987, Krylov2000, Kowalski2001}
The particularly challenging nature of \ce{F2} is present even at the HF level, where RHF vastly overestimates the binding energy (in common with most single bonds)\cite{Kowalski2001} and UHF predicts a completely unbound potential.\cite{Gordon1987, Laidig1987}
Moreover, the RHF solution provides a poor reference wave function at both equilibrium and dissociation geometries,\cite{Yang2013} posing further difficulties for post-HF methods.
For example, CCSD overbinds the molecule by almost a factor of two,\cite{Kowalski2001} while the ``gold standard'' CCSD(T) fails completely at large bond lengths.\cite{Kowalski2001}
Even conventional multireference methods struggle to describe the dissociation energy correctly, with full valence \CASSCF{14}{8} underestimating the potential well depth by around a factor of half.\cite{Gordon1987}
In contrast, CCSDT --- known to describe single bond breaking well\cite{Kowalski2001} --- provides a remarkably close approximation to the exact FCI potential energy surface computed by Bytautas and Ruedenberg,\cite{Bytautas2010} demonstrating the importance of triple excitations for capturing electron correlation.

Taking an active space for SCF metadynamics comprising the valence $3 \sigg$ bonding and $3 \sigu$ antibonding molecular orbitals (leaving the $\uppi$ orbitals frozen), we identify eight real UHF states in the dissociation limit that directly mirror the minimal basis states of \ce{H2}.\cite{Burton2016, Burton2018}
Folowing relaxation in the full orbital space, these states correspond to two spatially symmetry-broken UHF radical states (\ce{$^{\upharpoonleft}$F\bond{...}F^{$\downharpoonright$}} and \ce{$^{\downharpoonleft}$F\bond{...}F^{$\upharpoonright$}}), the bonding $\sigg^2$ and antibonding $\sigu^2$ RHF states, two non-bonding $\sigg \sigu$ UHF solutions, and two spatially symmetry-broken RHF states resembling the ionic configurations \ce{F^{+}\bond{...}F^{-}} and \ce{F^{-}\bond{...}F^{+}}, as shown in Fig.~\ref{fig:fluorine_cc-pvdz}.
Similarly to the multiple HF states of \ce{H2}, as the bond length is shortened we find two distinct Coulson--Fischer points involving the coalescence of the radical UHF and ionic RHF solutions with the $\sigg^2$ and $\sigu^2$ states respectively.
For shorter bond lengths, the corresponding h-RHF (dotted red) and h-UHF (dotted blue) solutions continue to exist with complex orbital coefficients.

We consider first the NOCI basis including only the $\sigg^2$ RHF ground state and the two radical sb-UHF states along with their h-UHF counterparts, which we denote NOCI(3).
Using this minimal NOCI basis yields a binding curve that closely matches the \CASSCF{14}{8}, suggesting that it captures a similar amount of the static correlation as \CASSCF{14}{8}, as shown in Fig.~\ref{fig:fluorine_cc-pvdz}.
Significantly, the variational flexibility offered by these three determinants appears sufficient to overcome the deficiency of the unbound UHF approximation, leading to a qualitatively correct bound potential while retaining size consistency in dissocation.
Like the full valence CASSCF, however, NOCI(3) underestimates the binding energy and overestimates the equilbrium bond length, indicating that NOCI(3) is still unable to capture any dynamic correlation that dominates at shorter geometries.

In contrast, when all eight h-HF solutions are included in the NOCI basis, denoted NOCI(8), the ground state energy is lowered further in comparison to the \CASSCF{14}{8}, particularly in the equilibrium region.
The additional multiple Hartree--Fock solutions appear to capture additional dynamic correlation at shorter bond lengths and, although NOCI(8) now overbinds the molecule, the absolute error is reduced in comparison to both NOCI(3) and CCSD.
Moreover, the equilibrium bond length predicted using NOCI(8) shows a very promising correspondence to the FCI and CCSDT, as shown in the bottom panel of Fig.~\ref{fig:fluorine_cc-pvdz}.
Overall we see that, even with such a small CI basis, NOCI provides a good reproduction of the relative binding curve of \ce{F2} across the full range of geometries, including where real HF solutions disappear and their h-HF counterparts become complex.
\subsection{Distortion of Cyclobutadiene}

\begin{figure}[bthp]
\centering
\includegraphics[scale=0.5, trim={20em 75ex 20em 75ex}, clip, keepaspectratio=true]{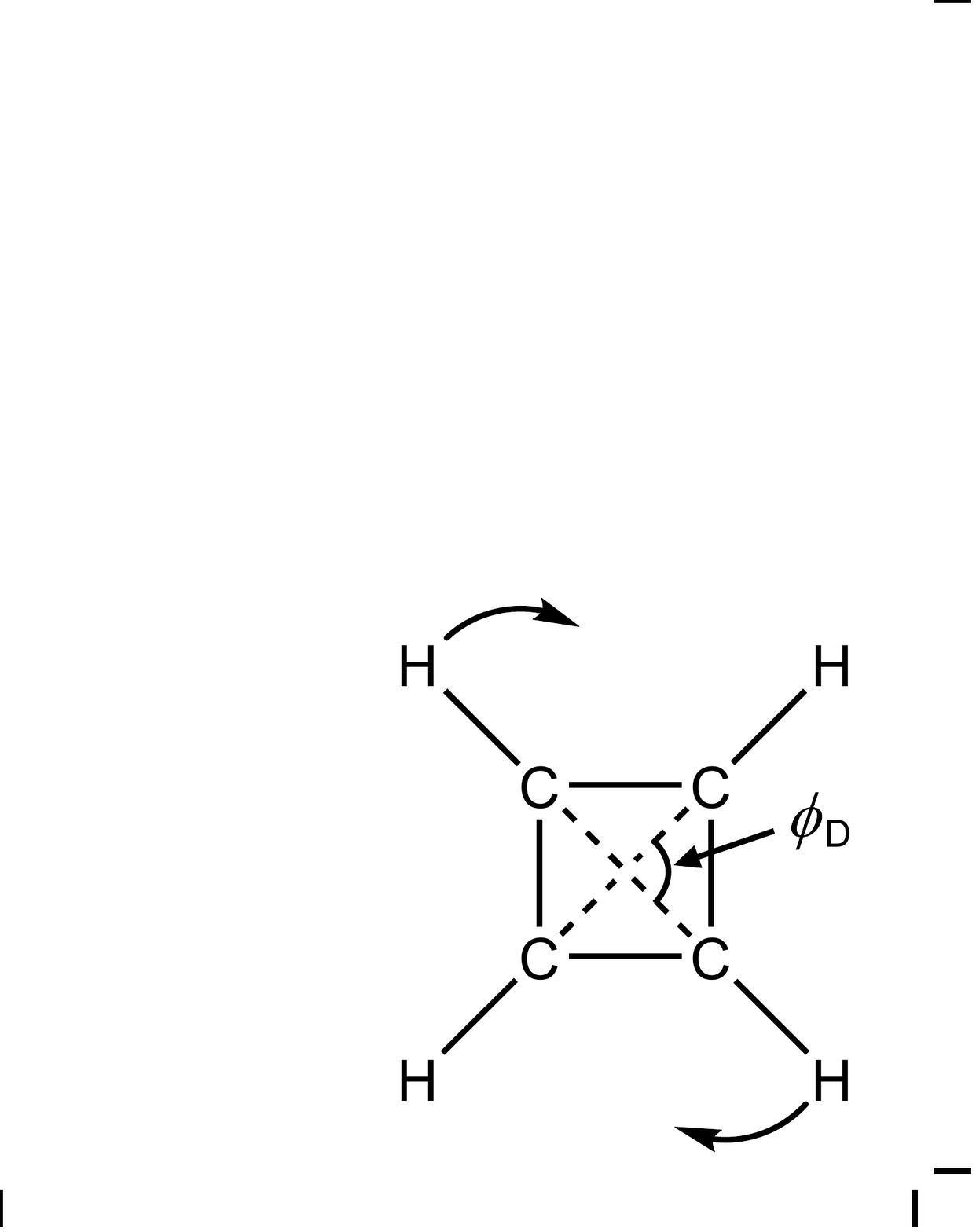}
\caption{Schematic demonstrating the distortion of cyclobutadiene used, where two corners of the square are rotated by the distortion angle $\DisAngle$ around the central $\mathcal{C}_{4}$ rotation axis.
For all distortion angles $\DisAngle$, the \ce{C-H} and diagonal \ce{C-C} distances are held constant.}
\label{fig:cyclobutadiene_scheme}
\end{figure}

\begin{figure}[t]
\centering
\includegraphics[width=0.5\textwidth, keepaspectratio=true]{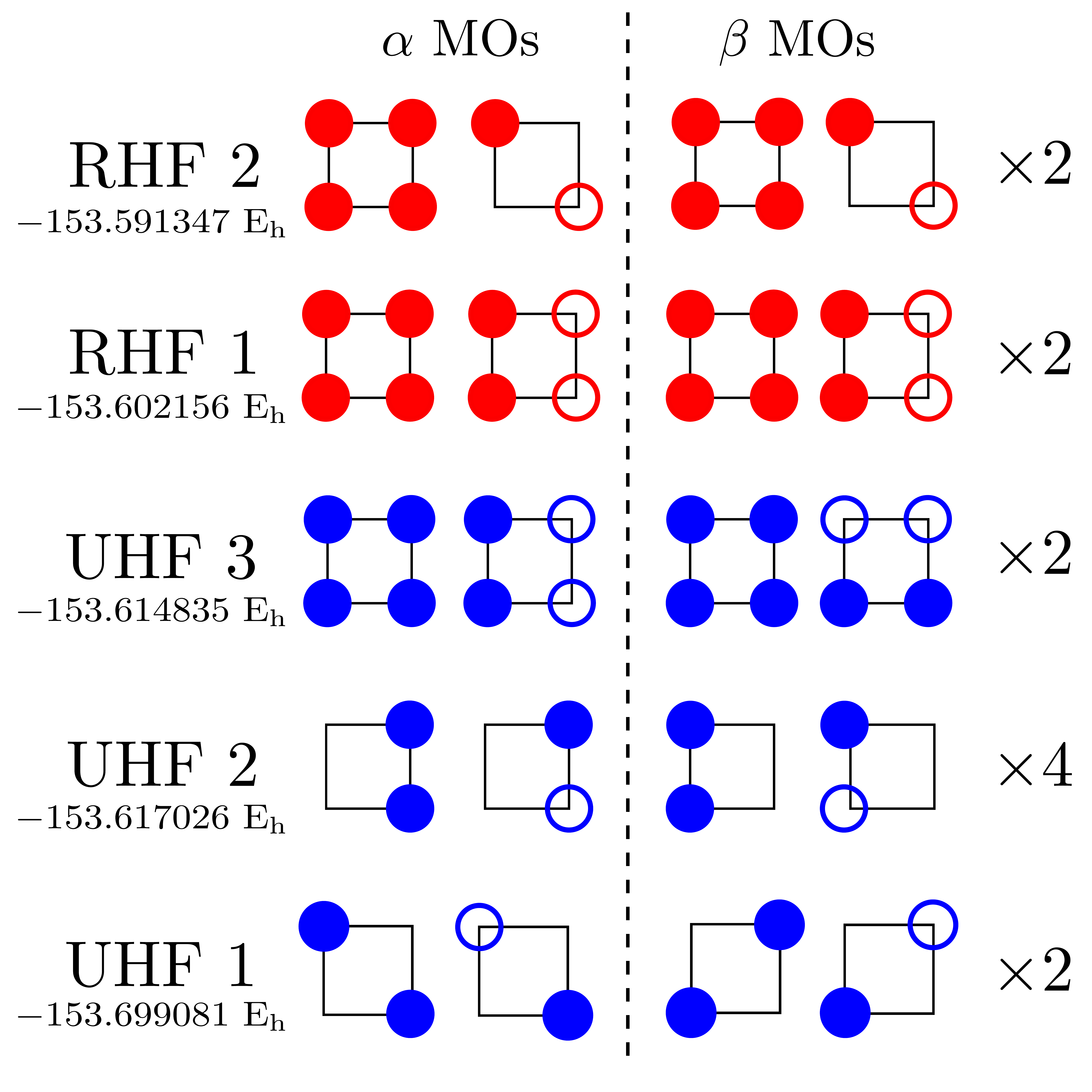}
\caption{Sketch of the $\alpha$ (left) and $\beta$ (right) occupied $\uppi$ molecular orbitals for the multiple RHF (red) and UHF (blue) states of cyclobutadiene, along with the HF energy and degeneracy of each state at the \DIVh{} geometry. 
Empty/filled circles (not to scale) indicate significant negative/positive contributions to the orbitals from the out-of-plane carbon p-orbitals.  
See main text for more details.}
\label{fig:cyclobutadiene_MO_sketch}
\end{figure}

\begin{figure*}[htp]
\centering
\includegraphics[width=\textwidth, keepaspectratio=true]{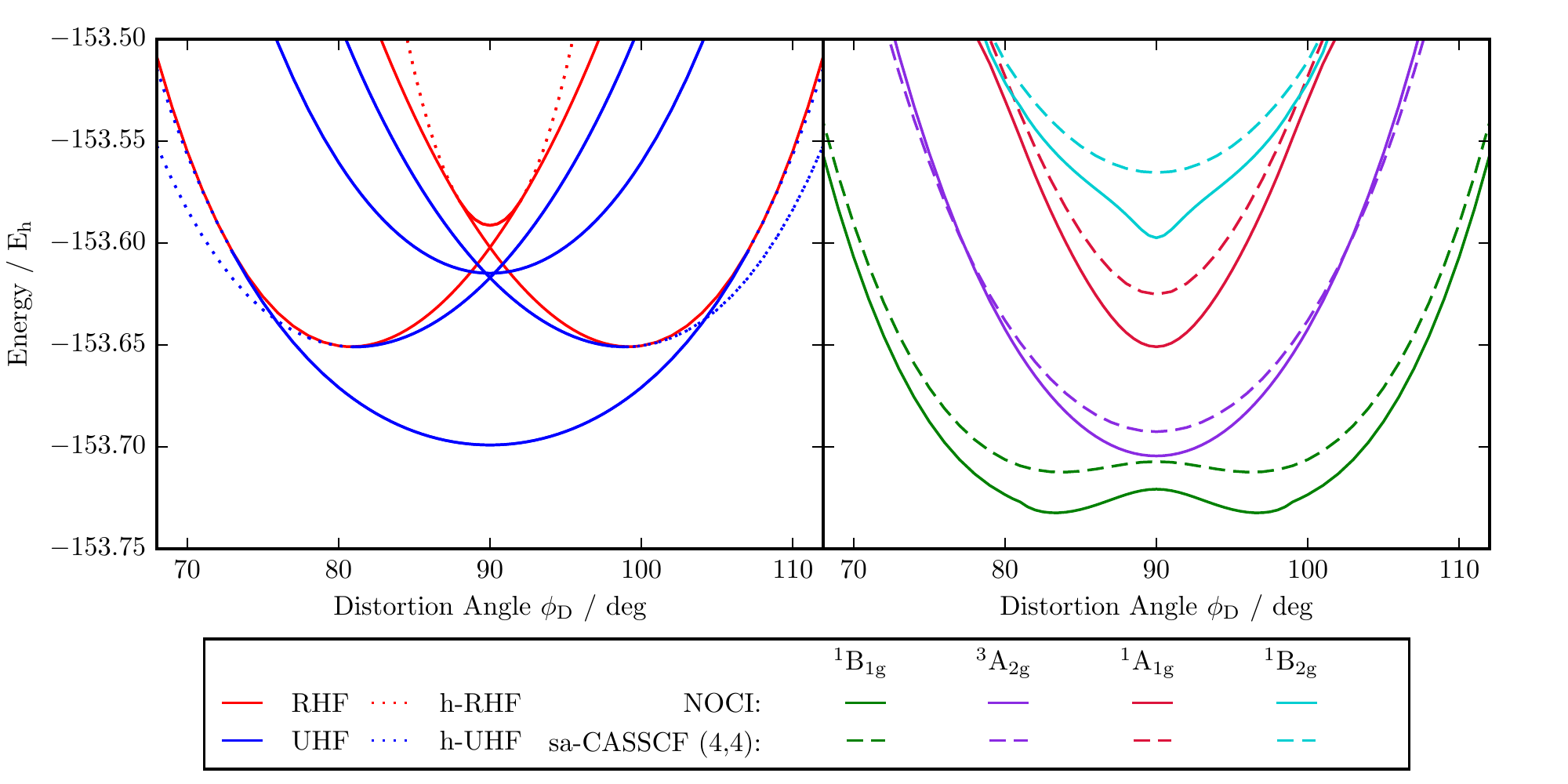}
\caption{\textit{Left}: Real RHF (solid red) and UHF (solid blue) states for the distortion of cyclobutadiene.
When real states disappear, their h-HF counterparts continue as complex h-RHF (dotted red) and h-UHF (dotted blue) solutions. Only the real part of any complex h-HF energies is plotted.\\
\textit{Right}: The four lowest energy NOCI states computed using these multiple h-HF solutions, compared to the sa-\CASSCF{4}{4} energies for the lowest four states.
Term symbols are given relative to the square \DIVh{} geometry.}
\label{fig:c4h4_absolute_cc-pvdz}
\end{figure*}

\begin{figure}[tbp]
\centering
\includegraphics[width=0.5\textwidth, keepaspectratio=true]{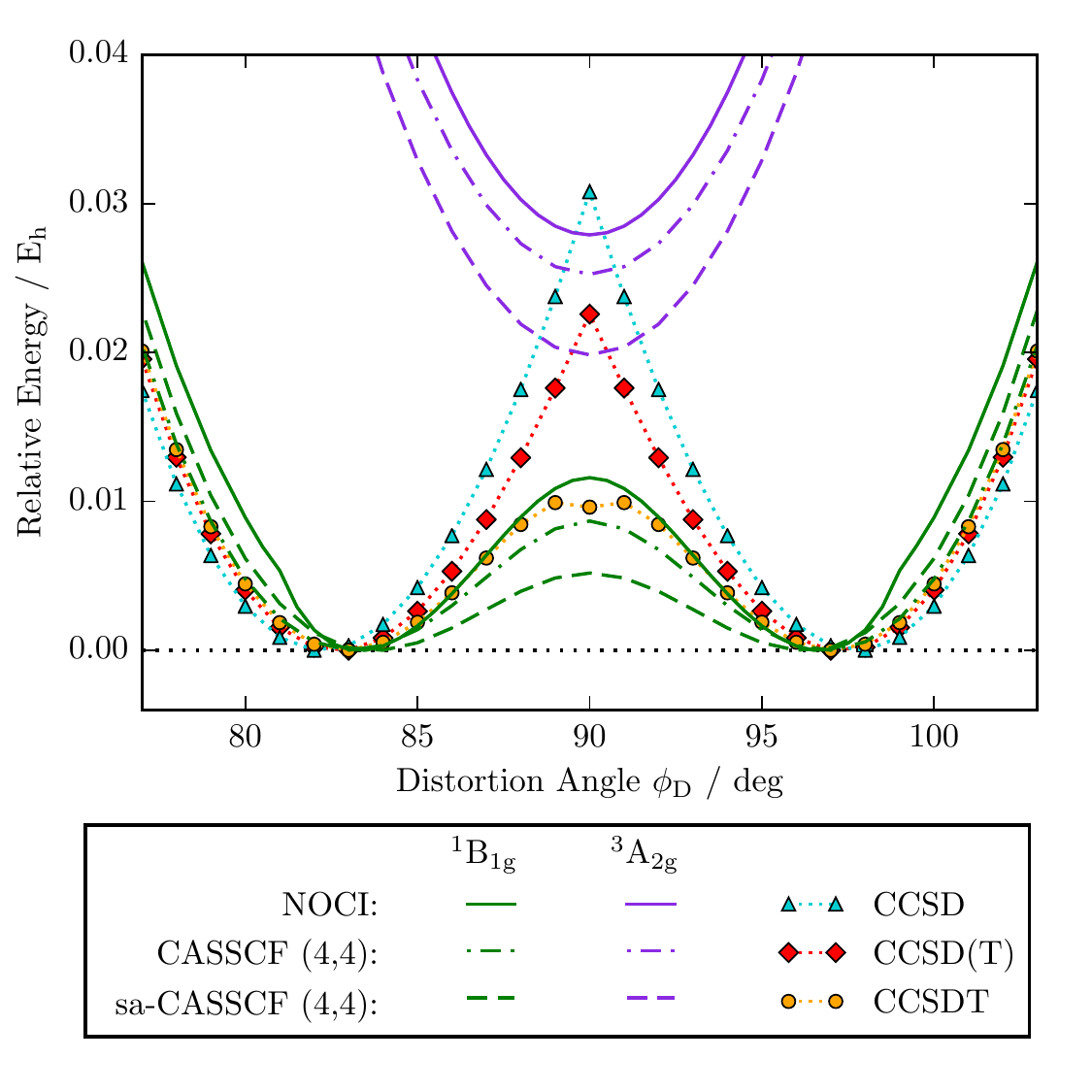}
\caption{Comparison of various methods for the relative autoisomerisation barrier and first excited state for cyclobutadiene.
Each curve is plotted relative to the ground state minimum energy for the corresponding method.
Term symbols are given relative to the square \DIVh{} geometry} 
\label{fig:c4h4_relative_cc-pvdz}
\end{figure}

Next we consider the cyclobutadiene molecule, which has long been of interest as an archetypal anti-aromatic and highly reactive system.\cite{Fratev1982, Kollmar1977, Borden1978, Buenker1968,Nakamura1989, Voorhis2000,Lyakh2011, Shen2008, Demel2008, Li2009, Deustua2017a, Shen2012,Musia2011, Yang2013, Rishi2017,  Levchenko2004, Sancho-Garca2000,  Carsky1988, Maksic2000, Eckert-Maksic2006, Saito2010, Li2009, Varras2018,Voter1986}
Correctly describing the ground and excited states of cyclobutadiene has formed the focus of extensive theoretical research.\cite{ Fratev1982, Kollmar1977, Borden1978, Buenker1968} 
In particular methods must balance the multireference nature of the square \DIVh{} geometry with the single-configurational character of the rectangular \DIIh{} energetic minimum.

At the \DIVh{} geometry, the highest occupied molecular orbital (HOMO) is a doubly-degenerate pair of singly occupied orbitals, leading to the $\uppi$-configuration $(\mathrm{a_{2u}})^2 (\mathrm{e_g})^2$.
This configuration results in a $\term{1}{B}{1g}$ ground state and a low-lying $\term{3}{A}{2g}$ first excited state, followed by two higher energy states of symmetry $\term{1}{A}{1g}$ and $\term{1}{B}{2g}$.\cite{Levchenko2004}
As the nuclear coordinates distort towards the rectangular geometry, the molecular symmetry drops from \DIVh{} to \DIIh{} and the $\uppi$-configuration becomes $(\mathrm{b_{1u}})^2 (\mathrm{b_{2g}})^2$.
The associated descent in symmetry of the ground state ($\term{1}{B}{1g}$ becomes $\term{1}{A}{g}$) and excited singlet state ($\term{1}{B}{2g}$ becomes $\term{1}{A}{g}$) leads to a second-order pseudo-Jahn--Teller effect favouring distortion towards the rectangular geometry.\cite{Buenker1968, Nakamura1989, Borden1978, Voorhis2000}

Significantly, while the ground-state RHF solution provides a suitable reference for the $\pointgroup{D}{2h}$ geometry, it becomes doubly-degenerate at the \DIVh transition state with only one of the degenerate HOMO orbitals being doubly-occupied in each degenerate RHF solution (sketched as \RHF{1} in Fig.~\ref{fig:cyclobutadiene_MO_sketch}).
As a result, single reference methods such as CCSD and CCSD(T) fail to provide even a qualitatively accurate description of the energy surface, with unphysical cusps propagated from the RHF description at the square geometry, as shown in Fig.~\ref{fig:c4h4_relative_cc-pvdz}.
Removing these cusps requires either the full inclusion of triple excitations (ie.~CCSDT),\cite{Musia2011, Shen2012, Deustua2017a} or multireference approaches such as multi-configurational SCF,\cite{Varras2018, Nakamura1989} generalised valence-bond theory,\cite{Voter1986, Carsky1988} or multireference CC.\cite{Lyakh2011, Shen2008, Li2009, Sancho-Garca2000, Balkova1994}

Starting from the rectangular geometry optimised at the RHF level (see \SI{}), we replicate the pseudo-Jahn--Teller distortion through the square geometry by simultaneously rotating two opposite corners around the central $\mathcal{C}_4$ rotation axis with the distortion angle $\DisAngle$, as illustrated in Fig.~\ref{fig:cyclobutadiene_scheme}.
For each distortion angle $\DisAngle$, the \ce{C-H} and diagonal \ce{C-C} distances, are held constant.
At the square \DIVh{} transition state geometry we define an active space for SCF metadynamics using the four $\uppi$ orbitals ($\mathrm{a_{2u}}$, $\mathrm{e_g}$ and $\mathrm{b_{2u}}$) from the ground state RHF solution and identify twelve low-energy real HF states.
After relaxation in the full orbital space, the degeneracies of these states (in order of ascending energy) are 2, 4, 2, 2 and 2. 
In what follows, we refer to the $n^{\text{th}}$-lowest RHF and UHF states at the square geometry using the notation ``\RHF{$n$}'' and ``\UHF{$n$}'' respectively.   

Inspecting the $\uppi$ orbitals for each solution --- sketched for one state of each degenerate set at the square geometry in Fig.~\ref{fig:cyclobutadiene_MO_sketch} --- we observe spatially symmetry-broken orbitals in the \UHF{1}, \UHF{2} and \RHF{2} solutions.
In contrast, the orbitals of the \UHF{3} states preserve spatial symmetry, representing the $(\mathrm{a_{2u}})^2 (\mathrm{e_g})^2$ configuration in which both orbitals in the degenerate $\mathrm{e_g}$ pair contain one electron each, while the \RHF{1} orbitals represent the same valence configuration but with only one of the degenerate $\mathrm{e_g}$ orbitals holding both electrons.
The degeneracies for each state can be deduced by considering the spatial and spin symmetries of the system.
As we move away from the square geometry, the degeneracy of both the \RHF{1} and \UHF{2} states is broken, splitting into lower (higher) energy states \RHF{1a} (\RHF{1b}) and \UHF{2a} (\UHF{2b}) with one- and two-fold degeneracies respectively.
The single degeneracy of the \RHF{1a} gtound state leads to the dominant single-reference character at the rectangular geometry.
In addition, each of the symmetry-broken solutions coalesces with the symmetry-pure \RHF{1a} state at a different Coulson--Fischer point, as show in the left panel of Fig.~\ref{fig:c4h4_absolute_cc-pvdz}.
For distortion angles further away from $90\degree$, the h-HF counterparts of the vanishing states continue to exist with complex orbital coefficients (dotted lines in Fig.~\ref{fig:c4h4_absolute_cc-pvdz}).


Using these HF solutions as a basis for NOCI, we recover continuous and smooth energies that are all variationally lower than their sa-\CASSCF{4}{4} counterparts (right panel in Fig.~\ref{fig:c4h4_absolute_cc-pvdz}).
Again, NOCI recovers the static correlation required to provide the correct qualitative description of the ground and excited states, while the individually optimised HF basis states capture additional dynamic correlation that quantitively improve the energy relative to sa-CASSCF.
Moreover, NOCI yields a smooth relative ground-state autoisomerisation barrier, shown in Fig.~\ref{fig:c4h4_relative_cc-pvdz}, with a close agreement to CCSDT.
In fact, even a state-specific \CASSCF{4}{4} ground-state calculation fails to reproduce the relative accuracy of NOCI.
We note, however, that the NOCI ground state exhibits an apparent bump at around $\DisAngle = 99 \degree$. This feature results from a complicated ``avoided crossing'' of the complex h-HF extension of the \RHF{2} state with another complex solution in the complex $\lambda$-plane, although a detailed investigation is beyond the scope of the current communication.
Crucially, the h-HF states used are consistent in the vicinity of this bump without any coalescence, and thus the NOCI energy remains both smooth and continuous.

\begin{table}[tpb]
\begin{ruledtabular}
\begin{tabular}{ld{4.6}d{2.6}d{2.6}d{2.6}}
			&	\head{$\Etot (\term{1}{B}{1g})$}	&	\head{$\term{3}{A}{2g}$} 	& \head{$\term{1}{A}{1g}$}	& \head{$\term{1}{B}{2g}$} 
	\Tstrut\Bstrut\\
	\hline 
	NOCI 						&	-153.72075		&	0.01630		&	0.06987		&	0.12331		
	\Tstrut\\
	CIS            				& 	-153.60216		&  -0.06703		&	0.01840		&	0.22862 
	\\
	\CASSCF{4}{4}				&	-153.70956		&	0.01657		&	\text{---}		&	\text{---}				
	\\
	sa-\CASSCF{4}{4}				& 	-153.70721		& 	0.01465		&	0.08204		&	0.14176 
	\\
	EOM-CCSD 					& 	-154.18974		&  -0.02062		&	0.24389		&	0.25355	 
	\\
	EOM-CCSDT 					& 	-154.23741 		&	0.00271		&	\text{---}		&	\text{---}	
	\\ \hline\Tstrut
	ex-FCI 						& 	-154.234(1)   	& 0.013(3)		& 0.058(2)		& 0.078(1)   
\end{tabular}
\end{ruledtabular}
\caption{Total energy $\Etot$ of the singlet ground state, and first three vertical excitation energies for cyclobutadiene.}
\label{tab:CyclobutadieneSqExEn}
\end{table}

Finally, we consider the vertical excitation energies for \DIVh{} cyclobutadiene. 
At this geometry, dynamic correlation is known to lower the singlet below the triplet state,\cite{Levchenko2004, Kollmar1977, Nakamura1989} leading to a violation of Hund's rules through the dynamic spin-polarisation effect.
In Table~\ref{tab:CyclobutadieneSqExEn} we compare the excitation energies calculated using NOCI to those computed using CIS, EOM-CCSD, EOM-CCSDT, both state-specific \CASSCF{4}{4} and sa-\CASSCF{4}{4}, and ex-FCI.
In sa-\CASSCF{4}{4}, we simultaneouly optimise the four lowest energy states, while in the state-specific variant we focus on only the lowest energy singlet and triplet states in separate calculations.
For the $(\term{1}{B}{1g} \rightarrow\ \term{3}{A}{2g})$ transition, NOCI matches the state-specific \CASSCF{4}{4} result, although both overestimate the excitation energy. 
In contrast, the uncorrelated CIS leads to an incorrect ordering of the singlet and triplet states, as does the absence of static correlation in EOM-CCSD.
For the higher energy singlet-singlet transitions, NOCI gives the closest estimate to the ex-FCI result in comparison to all the methods considered. 
In contrast, EOM-CCSD significantly overpredicts the excitation energies, while we were unable to converge EOM-CCSDT calculations for these transitions.
Overall, NOCI is able to match the performance of state-specific \CASSCF{4}{4} and outperform EOM-CCSD to provide accurate estimates of multireference excitation energies at a significantly lower cost.

\section{Concluding Remarks}

We have provided a general protocol for using multiple h-HF solutions to construct NOCI wave functions that accurately describe multireference ground and excited energies across all molecular geometries.
In particular, we have presented two key algorithmic advances.
Firstly, we described an active space SCF metadynamics approach that allows a chemically relevant subset of real HF solutions to be identified.
Secondly, to handle any real HF solution that disappears and routinely extend it as a complex h-HF state --- and thus ensure smooth NOCI energies --- we have presented an approach that exploits the complex topology of multiple h-HF states.
We have shown that this combined approach can capture correlation to a similar, if not better, accuracy than CASSCF using an equivalent active space, and can provide accurate multireference excitation energies with remarkably small NOCI expansions.
Crucially, this NOCI approach is both systematically improvable and scales very favourably with the system size.

The applications of using fully optimised multiple HF solutions as a basis for NOCI still remain relatively unexplored, and we believe our approach holds great potential for larger systems that are out of reach for conventional multireference techniques.
In particular, the inclusion of symmetry-broken solutions that qualitatively represent physical states can provide chemical insight into complicated processes such as electron transfer.\cite{Jensen2018}
Moreover, retaining the use of fully optimised HF solutions in the NOCI wave function paves the way for the derivation of molecular gradients and, although NOCI captures mainly static correlation, it also provides a foundation for post-NOCI correlation techniques such as NOCI-MP2.\cite{Yost2013,Yost2016}
Ultimately, by combining SCF metadynamics and h-HF theory to create a single systematic and routine approach, we have laid the foundations for using NOCI to describe multireference ground and excited states in general.

\section*{Acknowledgements}

H.G.A.B.~thanks the Cambridge Trust for a studentship and Q-Chem for supporting code development of this work through a summer internship.
A.J.W.T.~thanks the Royal Society for a University Research Fellowship (UF110161). 

\bibliography{HoloActiveSpaces,HoloActiveSpace-control}

\end{document}